\documentstyle[12pt]{article}
\begin{document}
\begin{center}
{\large \bf Non-Kramers degeneracy and oscillatory tunnel
splittings in the biaxial Spin System}\\

Degang Zhang$^{1, 2, 3}$ and Bambi Hu$^{2, 1}$\\

{$^{1}$Department of Physics, University of Houston, Houston, TX
77204, USA\\ $^{2}$Department of Physics and Centre for Nonlinear
Studies, Hong Kong Baptist University, Kowloon Tong, Hong Kong,
China\\
 $^{3}$Institute of Solid State Physics, Sichuan Normal
University, Chengdu 610068, China}
\end{center}

\begin{abstract}

We have investigated analytically quantum tunneling of large spin
in the biaxial spin system with the magnetic field applied along
the hard and medium anisotropy axes by using a purely
quantum-mechanical approach. When the magnetic field parallels the
hard axis, the tunnel splittings of all the energy level pairs are
oscillatory as a function of the magnetic field. The quenching
points are completely determined by the coexistence of solutions
of the Ince's equation. When the magnetic field points the medium
axis, the tunnel splitting oscillations disappear due to no
coexistence of solutions. These results coincide the recent
experimental observations in the nanomagnet Fe$_8$.

\end{abstract}


In recent years much attention has been paid to quantum tunneling
of magnetization (large spin) in nanomagnets, both from experiment
and from theory [1]. Several magnetic particles have been
identified as promising candidates for the observation of such
macroscopic quantum phenomena, where the magnetization (or the
N\'{e}el vector) tunnels from one potential minimum to another
one. The excellent examples that have widely studied are the
molecular nanomagnets Fe$_8$ [2-5] and Mn$_{12}$ [6-8], which have
the well-defined structures and magnetic properties. On the one
hand, these phenomena are very interesting from a fundamental
point of view because they extend our understanding of the
transition between quantum and classical behavior. On the other
hand, tunneling of the magnetization changes the magnetic
properties of the nanomagnets, which has the potential application
for the data storage technology, e.g., making qubits - the
elements of quantum computers.

Recently Wernsdorfer and Sessoli [9] have observed  a novel
phenomenon, i.e. oscillations of tunnel splittings of the ground
and excited states in the nanomagnet Fe$_8$ described well by the
spin Hamiltonian [2-5] $${\mathcal{H}}=AS^2_x-BS^2_y-g\mu_B{\bf
S\cdot H},\eqno{(1)}$$ where ${\bf S}$ is a spin operator, ${\bf
H}$ is the magnetic field applied in the x-z plane, the spin
quantum number $s=10$, $A\approx 0.092$K, $B\approx 0.229$K,
$g\approx 2$ is the $g$-factor, $\mu_B$ is the Bohr magneton. The
zero-field Hamiltonian has a biaxial symmetry with hard, easy and
medium axes along x, y and z, respectively. When ${\bf H }$
rotates from ${\bf x}$ to ${\bf z}$ direction, the oscillations of
tunnel splittings of all the level pairs gradually disappear.

In fact, oscillation of the ground state tunnel splitting $\Delta
E_s$ of the model (1) with the magnetic field $H_x$ (${\bf H
\parallel x}$) was predicted by using instanton technique [10].
The tunneling of spin is quenched when
$$h=h^*(1-\frac{n}{s}-\frac{1}{2s}), ~~~~~n=0, 1, \cdots,
2s-1.\eqno{(2)}$$ Here, $h=\frac{H_x}{H_c}$,
${H_c}=\frac{2s(A+B)}{g\mu_B}$ is the critical field at which the
energy barrier vanishes, $-h^*<h<h^*=\sqrt{\frac{A}{A+B}}$, and
$s$ is an integer or half odd integer. This kind of topological
quenching is the result of quantum interference of different
instanton paths within the context of macroscopic quantum
tunneling [11, 12], and need not be related to Kramers's
degeneracy. Up to the first order of the rate $\frac{B}{A+B}$, the
formula (2) was rederived by quantum-mechanical perturbation
theory [13]. Very recently, Garg  extended his previous work [10]
to the excited states by using a discrete
Wentzel-Kramers-Brillouin approach [14]. To order $s^{-1}$, the
quenching points for the excited state pairs are the same as those
for the ground state pair. With increasing the magnetic field, the
quenching points gradually decrease and finally disappear. These
results were also obtained by the potential field description of
spin systems with exact spin-coordinate correspondence [15] and
the instant technique [16, 17]. Because quantum tunneling of spin
in the nanomagnets was observed at very low temperatures, the
quantization of spin levels becomes very important to explain well
the experiments. In essence, the energy spectrum of the spin
systems can help us to understand  the mechanism of spin tunneling
thoroughly. In this paper, we have  analytically diagonalized the
Hamiltonian (1) in the framework of the Schr\"{o}dinger's picture
of quantum mechanics. The energy spectrum of the spin model is
obtained in the large-$s$ limit. It is clearly shown that when
${\bf H \parallel x}$, the tunnel splittings of the ground and
excited state pairs are oscillatory as a function of $H_x$ and the
quenching points agree with the numerical simulation of model (1).
When ${\bf H \parallel z}$, the tunnel splitting oscillations of
all the energy level pairs disappear, which coincided with those
obtained by the phase space path integral [18]. These phenomena
have also been observed in the experiments [9, 19].

Let $E$ and $\Phi_m$ be the eigenenergies and eigenstates of
$\mathcal{H}$, respectively, then the eigenvalue equation in the
basis $|s, m>_z$ reads $$u_{m-1}\Phi_{m-2}+u_{m+1}\Phi_{m+2}
-t_{m-\frac{1}{2}}\Phi_{m-1}- t_{m+\frac{1}{2}}\Phi_{m+1}$$
$$+\{-E+\frac{1}{2}(A-B)[s(s+1)-m^2]-g\mu_BH_zm\}\Phi_m=0,
\eqno{(3)}$$ where $$u_{m\pm
1}=\frac{1}{4}(A+B)\sqrt{[s(s+1)-(m\pm 1)^2]^2-(m\pm 1)^2},$$
$$t_{m\pm \frac{1}{2}}=\frac{1}{2}g\mu_BH_x\sqrt{s(s+1)-
 (m-\frac{1}{2})^2+\frac{1}{4}}$$. Obviously, it is
very difficult to strictly solve the equation (3) for arbitrary
$s$. However, in the large-$s$ limit, Eq. (3) becomes [20]
$$(1-x^2)\frac{d^2\Phi}{dx^2}-2x\frac{d\Phi}{dx}+[-\frac{E}{A+B}-\frac{1}{4}
-\frac{1}{4}\frac{1}{1-x^2}$$
$$+\frac{As(s+1)}{A+B}(1-x^2)-\frac{g\mu_BH_z\sqrt{s(s+1)}}{A+B}x$$
$$-\frac{g\mu_BH_x\sqrt{s(s+1)}}{A+B}\sqrt{1-x^2}]\Phi=0.\eqno{(4)}$$
Here, $x=\frac{m}{\sqrt{s(s+1)}}$ and only the leading terms are
remained, i.e. $0(s^{-1})$. In deriving Eq. (4), we have used

$$u_{m\pm 1}\Phi_{m\pm 2}=u_{m\pm 1}[\Phi_{m\pm 1} \pm
\frac{\Phi_{m\pm 1}^\prime}{\sqrt{s(s+1)}}+\frac{\Phi_{m\pm
1}^{\prime\prime}}{2s(s+1)}+\cdots]$$ $$=u\Phi
\pm\frac{1}{\sqrt{s(s+1)}}\frac{d}{d x}(u\Phi)+ \frac{1}{2s(s+1)}
\frac{d^2}{d x^2}(u\Phi)+\cdots$$ $$\pm
\frac{1}{\sqrt{s(s+1)}}[u\frac{d\Phi}{d x}\pm
\frac{1}{\sqrt{s(s+1)}}\frac{d}{d x}(u\frac{d\Phi}{d x})+\cdots]$$
$$+ \frac{u}{2s(s+1)}\frac{d^2\Phi}{d x^2}+\cdots,$$ $$t_{m\pm
\frac{1}{2}}\Phi_{m+1}=t_{m\pm \frac{1}{2}}[\Phi_{m\pm
\frac{1}{2}} \pm \frac{\Phi_{m\pm
\frac{1}{2}}^\prime}{2\sqrt{s(s+1)}}+\cdots] $$ $$=t\Phi\pm
\frac{1}{2\sqrt{s(s+1)}}\frac{d}{dx}(t\Phi)+\cdots$$ $$\pm
\frac{t}{2\sqrt{s(s+1)}}\frac{d\Phi}{dx}+\cdots,$$ where $$
\begin{array}{ccc}
 u&=& \frac{1}{4}(A+B)[\sqrt{s^2(s+1)^2(1-x^2)^2-s(s+1)x^2}\\
 &\approx &
\frac{1}{4}(A+B)[s(s+1)(1-x^2)-x^2/[2(1-x^2)]\end{array}$$ and
$t\approx \frac{1}{2}g\mu_BH_x\sqrt{s(s+1)}\sqrt{1-x^2}$ for large
$s$.

Taking the transformations: $\Phi=(1-x^2)^{-\frac{1}{4}}y(x)$ and
$x={\rm sin}(2t)$, and substituting them into Eq. (4), we finally
obtain the Hill's equation [21]

$$\frac{d^2y}{dt^2}+[\Lambda+(a-b){\rm cos}(2t)+(c-b){\rm sin}(2t)
+\frac{b^2}{8}{\rm cos}(4t)]y=0, \eqno{(5)}$$ where
$$\Lambda=\frac{-4E+2As(s+1)}{A+B},
a=b-\frac{4g\mu_BH_x\sqrt{s(s+1)}}{A+B},$$
$$b=\pm4\sqrt{\frac{As(s+1)}{A+B}},
c=b-\frac{4g\mu_BH_z\sqrt{s(s+1)}}{A+B}. \eqno{(6)}$$ Up to now,
we have mapped the spin problem (1) onto a particle problem (5).
The energy spectrum of $\mathcal{H}$ is completely determined by
the characteristic levels of the Hill's equation.  We note that
when $A=0$ and ${\bf H }=0$, $\Delta=m^2$. So
$E=-B(\frac{m}{2})^2$ are nothing but the eigenvalues of the
Hamiltonian (1) with integer $s$ for even $m$ or with half odd
integer $s$ for odd $m$.

For Eq. (5), there exist two monotonically increasing sequences of
real numbers $a_0, a_{2i}, b_{2i}, a_{2i-1}^\prime$ and
$b_{2i-1}^\prime (i=1, 2, \cdots)$ such that Eq. (5) has a
solution with period $\pi$ if and only if $\Lambda=a_0, a_{2i}$ or
$b_{2i}$, and a solution with period $2\pi$ if and only if
$\Lambda=a_{2i-1}^\prime$ or $b_{2i-1}^\prime$. The $a_0, a_{2i},
b_{2i}, a_{2i-1}^\prime$ and $b_{2i-1}^\prime$ satisfy
inequalities: $$a_0<b_1^\prime\leq a_1^\prime<b_2\leq a_2<
b_3^\prime\leq a_3^\prime<b_4\leq a_4<\cdots. \eqno{(7)}$$
According to the relation between $E$ and $\Lambda$ in Eq. (6), it
is easy to see that the ground state of $\mathcal{H}$ corresponds
to the allowed highest characteristic level of the Hill's equation
with period $\pi$ or $2\pi$, depending on integer or half odd
integer spin $s$. The lower the characteristic level of Eq. (5)
is, the higher the associated eigenstate of the Hamiltonian (1)
is. The tunnel splitting of the characteristic level pair of Eq.
(5) with period $\pi$ is $$\Delta \Lambda_{2i}=a_{2i}-b_{2i},
\eqno{(8)}$$ and the tunnel splitting of the characteristic level
pair of Eq. (5) with period $2\pi$ is $$\Delta
\Lambda_{2i-1}^\prime=a_{2i-1}^\prime-b_{2i-1}^\prime.
\eqno{(9)}$$ The tunnel splitting of the energy level pair of the
Hamiltonian (1) can be evaluated by the tunnel splitting of its
associated characteristic level pair using Eq.(6). We note that
when two of three parameters $A$, $H_x$ and $H_z$ vanish, the
Hill's equation (5) becomes the well-known Mathieu equation [22].
So the tunnel splittings of all the energy level pairs of the
Hamiltonian (1) are monotonous increasing rather than oscillatory
in the three cases [20]. For arbitrary parameters $A$, $B$, $H_x$
and $H_z$, it is difficult to solve analytically Eq. (5). Here we
only consider the following two special cases.

(i) $H_z=0$ (i.e. $c=b)$. In this case, Eq. (5) reduces to the
Ince's equation, which has been studied in a set of literature due
to its physically basic importance [21]. To observe the
oscillations of tunnel splittings of model (1), we must find some
vanishing points at which its eigenstates are degenerate, i.e.
$a_{2i}=b_{2i}$ or $a_{2i-1}^\prime=b_{2i-1}^\prime$. This is
equivalent to find the coexistence of solutions of the Ince's
equation, which means that there exist two linearly independent
solutions (one even and one odd) with period $\pi$ or $2\pi$. Very
fortunately, due to the positive coefficient of the last term
${\rm cos}(4t)$, the Ince's equation has the coexistence of
solutions [21] with period $\pi$ when $$a=-2nb\eqno{(10)}$$ and
with period $2\pi$ when $$a=-(2n-1)b,\eqno{(11)}$$ where $n=0, \pm
1, \pm 2, \cdots$. Obviously, the quenching points of tunnel
splittings of the characteristic level pairs with period $\pi$ are
exactly shifted by half a period relative to those with period
$2\pi$. Under the conditions (10) and (11), the Ince's equation
becomes the Whittaker equation [21] $$\frac{d^2y}{dt^2}+[\Lambda-p
b{\rm cos}(2t)+\frac{b^2}{8}{\rm cos}(4t)]y=0. \eqno{(12)}$$ Here,
$p=2n+1$ or $2n$ for Eq. (10) or (11), respectively. For Eq. (12),
when $b\rightarrow 0$, then $a_0\rightarrow 0$, $a_{2i}$ and
$b_{2i}\rightarrow (2i)^2$, and $a_{2i-1}^\prime$ and
$b_{2i-1}^\prime \rightarrow (2i-1)^2$ [23].

{\it Integer spin s}. Due to Eq. (6), the eigenstates of the
Hamiltonian (1) correspond to the characteristic levels of the
Ince's equation with period $\pi$, i.e.
$E_0^a=-\frac{1}{4}(A+B)a_0+E_0,
E_i^a=-\frac{1}{4}(A+B)a_{2i}+E_0$ and
$E_i^b=-\frac{1}{4}(A+B)b_{2i}+E_0, E_0=\frac{1}{2}As(s+1), i=1,
2, \cdots, s$. So the tunnel splitting of the energy level pair
$(E_i^a, E_i^b)$ is $$\Delta
E_i=E_i^b-E_i^a=\frac{1}{4}(A+B)\Delta\Lambda_{2i}.\eqno{(13)}$$
From Eq. (10), we have $\Delta E_i=0$ when
$$H_x=\frac{(2n+1)\sqrt{A(A+B)}}{g\mu_B}. \eqno{(14)}$$ The
existence of the quenching fields (14) clearly shows that the
tunnel splittings of all the energy levels of $\mathcal{H}$ are
oscillatory as a function of $H_x$ and the period of oscillations
$\Delta H=\frac{2\sqrt{A(A+B)}}{g\mu_B}$. This coincides with
Garg's result (2) [10, 14]. For Eq. (12), when $|p|=2l+1$, then
the even intervals of instability on the $\Lambda$ axis disappear
with at most $l+1$ exceptions [21]. In other words, the
characteristic values $a_0, a_{2i}$ and $b_{2i}$ satisfy
$$a_0<b_2<a_2<\cdots<b_{2l}<a_{2l}$$
$$<b_{2(l+1)}=a_{2(l+1)}<\cdots< b_{2s}=a_{2s}. \eqno{(15)}$$ This
means that the tunnel splittings of the $l$ highest excited state
pairs of the Hamiltonian (1) do not vanish while those of the
other $s-l$ energy level pairs vanish. With increasing the
magnetic field $H_x$ (i.e. $|p|$), the quenching points gradually
decrease and finally disappear when $l\geq s$. The configuration
of the quenching points agree with that from the numerical
simulation of the Hamiltonian (1) (see Fig. 1).

{\it Half odd integer spin s}. The eigenstates of the Hamiltonian
(1) correspond to the characteristic levels of Eq. (12) with
period $2\pi$, i.e.
$E_{i-\frac{1}{2}}^{a^\prime}=-\frac{1}{4}(A+B)a_{2i-1}^\prime+E_0$
and
$E_{i-\frac{1}{2}}^{b^\prime}=-\frac{1}{4}(A+B)b_{2i-1}^\prime+E_0,
i=1, 2, \cdots, s+\frac{1}{2}$. The tunnel splitting of the energy
level pair $(E_{i-\frac{1}{2}}^{a^\prime},
E_{i-\frac{1}{2}}^{b^\prime})$ reads $$\Delta
E_{i-\frac{1}{2}}^{\prime}=E_{i-\frac{1}{2}}^{b^\prime}-
E_{i-\frac{1}{2}}^{a^\prime}=\frac{1}{4}(A+B)\Delta
\Lambda_{2i-1}^\prime.\eqno{(16)}$$ According to Eq. (11), we
obtain $\Delta E_{i-\frac{1}{2}}^\prime=0$ when
$$H_x^\prime=\frac{2n\sqrt{A(A+B)}}{g\mu_B}.\eqno{(17)}$$
Obviously, the period of oscillations of tunnel splittings for all
the energy level pairs $(E_{i-\frac{1}{2}}^{a^\prime},
E_{i-\frac{1}{2}}^{b^\prime})$ is $\Delta
H^\prime=\frac{2\sqrt{A(A+B)}}{g\mu_B}$, which is the same with
that of the energy level pairs $(E_i^a, E_i^b)$. However, the
quenching points for half odd integer spin $s$ are shifted half a
period (i. e.$ \frac{\sqrt{A(A+B)}}{g\mu_B})$ relative to those
for integer spin $s$. For Eq. (12), if $|p|=2l$, then at most
$l+1$ odd interval of instability on the $\Lambda$ axis remain,
 i. e.
$$b_1^\prime<a_1^\prime<b_3^\prime<a_3^\prime<\cdots<b_{2l-1}^\prime
<a_{2l-1}^\prime$$
$$<b_{2l+1}^\prime=a_{2l+1}^\prime<\cdots<
b_{2s}^\prime=a_{2s}^\prime. \eqno{(18)}$$ It is easy to see that
the tunnel splittings of the $s-l+\frac{1}{2}$ lowest energy level
pairs vanish, but the tunnel splittings of the other $l$ highest
energy level pairs do not. When $l>s-\frac{1}{2}$, there do not
exist the quenching points. These results also coincide with those
obtained by the discrete WKB approach [14].

(ii) $H_x=0$ (i.e. $a=b$). Let $t\rightarrow t+\frac{\pi}{4}$,
then Eq. (5) becomes the Ince's equation
$$\frac{d^2y}{dt^2}+[\Lambda+(c-b){\rm cos}(2t)-\frac{b^2}{8}{\rm
cos}(4t)]y=0. \eqno{(19)}$$ Because the coefficient of the last
term ${\rm cos}(4t)$ is negative, Eq. (19) does not possess the
coexistence of solutions, i.e. two linearly independent solutions
[21]. This means that the sign of equality in the inequalities (7)
cannot hold and each energy level of $\mathcal{H}$ is singlet in
the parameter space. So the tunnel splittings of all the level
pairs are not oscillatory with $H_z$, which coincide the
experiment [9, 19].

In conclusion, we  studied quantum tunneling of large spin in the
biaxial spin systems by using quantum mechanics. The energy
spectrum of the spin model (1) is obtained by solving  Hill's
equation (5), which is derived directly from the eigenvalue
equation of the spin problem in the large-$s$ limit. It is
surprising that when ${\bf H \parallel x}$, the vanishing points
of the tunnel splittings of all the energy level pairs obtained
here by the coexistence of solutions of the Ince's equation
coincide those given by the WKB method [10, 14] and other
approaches [15-17]. However, our approach is a natural way of
explaining the oscillations of tunnel splitting in the biaxial
spin system, which is also applied in other spin Hamiltonians.

This work was supported in part by grants from the Hong Kong
Research Grants Council (RGC), the Hong Kong Baptist University
Faculty Research Grant (FRG), the Sichuan Youth Science and
Technology Foundation and the NSF of the Sichuan Educational
Commission .

Fig. 1 The exact quenching points of all the level pairs of the
Hamiltonian (1) with $s=10, A=0.092K, B=0.229K,$ and $H_z=0$.


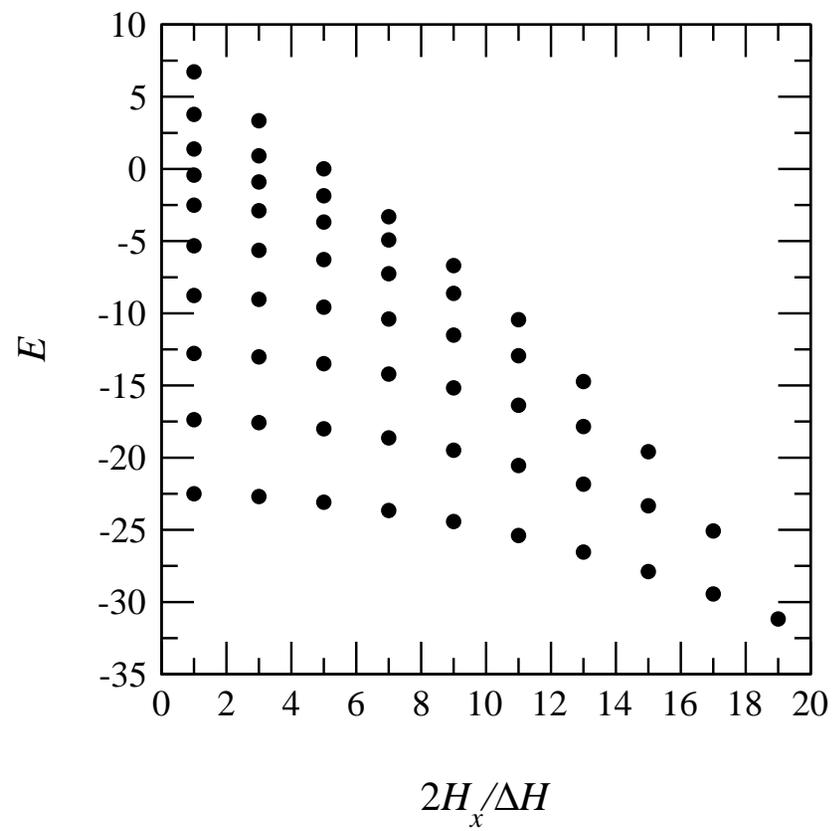
\end{document}